\documentclass[12pt]{article}
\makeatletter \@addtoreset{equation}{section} \makeatother
\renewcommand{\theequation}{\thesection.\arabic{equation}}
\newcommand{\ba}{\begin{array}}
\newcommand{\ea}{\end{array}}
\newcommand{\beq}{\begin{equation}}
\newcommand{\eeq}{\end{equation}}
\newcommand{\bea}{\begin{eqnarray}}
\newcommand{\eea}{\end{eqnarray}}

\def\bce{\begin{center}}
\def\ece{\end{center}}

\def\nonu{\nonumber}

\def\pa{\partial}

\def\be{\beta}

\def\ep{\epsilon}

\def\la{\lambda}

\def\eps6{{\displaystyle \mathop{\epsilon}^{6}}{}}
\def\g6{{\displaystyle \mathop{g}^{6}}{}}

\def\nab6{{\displaystyle \mathop{\nabla}^{6}}{}}

\def\0{{\sst{(0)}}}
\def\1{{\sst{(1)}}}
\def\2{{\sst{(2)}}}
\def\3{{\sst{(3)}}}
\def\4{{\sst{(4)}}}
\def\5{{\sst{(5)}}}
\def\6{{\sst{(6)}}}
\def\7{{\sst{(7)}}}
\def\8{{\sst{(8)}}}

\def\ba{\begin{array}}
\def\ea{\end{array}}
\def\beq{\begin{equation}}
\def\eeq{\end{equation}}
\def\be{\begin{equation}}
\def\ee{\end{equation}}

\def\la{\lambda}
\def\eps{\epsilon}

\def\ba{\begin{array}}
\def\ea{\end{array}}
\def\beq{\begin{equation}}
\def\eeq{\end{equation}}
\def\be{\begin{equation}}
\def\ee{\end{equation}}

\def\la{\lambda}
\def\eps{\epsilon}

\def\eps6{{\displaystyle \mathop{\epsilon}^{6}}{}}

\def\nab6{{\displaystyle \mathop{\nabla}^{6}}{}}

\newcommand{\bean}{\begin{eqnarray*}}
\newcommand{\eean}{\end{eqnarray*}}

\begin{document}
\thispagestyle{empty} \addtocounter{page}{-1}
   \begin{flushright}
\end{flushright}

\vspace*{1.3cm}
  
\centerline{ \Large \bf   
The Primary  Spin-4 Casimir Operators } 
\centerline{ \Large \bf 
in the  Holographic $SO(N)$ Coset Minimal Models}
\vspace*{1.5cm}
\centerline{{\bf Changhyun Ahn 
}} 
\vspace*{1.0cm} 
\centerline{\it 
Department of Physics, Kyungpook National University, Taegu
702-701, Korea} 
\vspace*{0.8cm} 
\centerline{\tt ahn@knu.ac.kr 
} 
\vskip2cm

\centerline{\bf Abstract}
\vspace*{0.5cm}

Starting from $SO(N)$ current algebra, we construct two lowest primary
higher spin-$4$ Casimir
operators which are quartic in spin-$1$ fields. 
For $N$ is odd, one of them corresponds to the
current in the $WB_{\frac{N-1}{2}}$ minimal model. 
For $N$ is even, the other  corresponds to the current   
in the $WD_{\frac{N}{2}}$ minimal model.  
These primary higher spin currents, the generators of wedge subalgebra, 
are obtained from the operator product 
expansion of fermionic (or bosonic) primary spin-$\frac{N}{2}$ field with
itself in each minimal model respectively.
We obtain, indirectly, 
the three-point functions with two real scalars, in the large $N$
't Hooft limit, for all values of the 't Hooft coupling which should be dual to the
three-point functions in the higher spin $AdS_3$ gravity with matter.  

\baselineskip=18pt
\newpage
\renewcommand{\theequation}
{\arabic{section}\mbox{.}\arabic{equation}}

\section{Introduction}

Gaberdiel and Gopakumar have conjectured in \cite{GG} 
that the large $N$ 't Hooft limit  of the 
$WA_{N-1}=W_N$ minimal model \cite{BBSS1} is dual to 
a particular $AdS_3$ higher spin theory of Vasiliev  \cite{PV,PV1,Vasiliev}. 
The boundary theory is an $A_{N-1}$ coset  model  
which has a higher spin $WA_{N-1}$ symmetry generated by currents of 
spins $s=2, 3, \cdots, N$ \cite{FL}. See the work of \cite{BS} 
for the $W$ symmetry in two-dimensional conformal field theory.
The theory is labeled by two positive integers $(N, k)$
where $k$ is the level of the current algebra and the 
't Hooft coupling $\lambda=\frac{N}{N+k}$ is fixed in the large
$N$ 't Hooft limit.
The bulk theory has an infinite tower of massless fields with spins
$s=2, 3, \cdots $  coupled to two complex scalars. 
The higher spin Lie algebra
describes interactions between the higher spin fields and the scalars.
The scalars  have equal mass  determined  by the algebra, $M^2 =
-(1-\lambda^2)$. The quantization
with opposite boundary conditions leads to 
their conformal dimensions $h_{\pm}= \frac{1}{2}(1 \pm \lambda)$.

The partition
function of the $WA_{N-1}$ minimal model was obtained in \cite{GGHR}. 
Since certain states become
null and decouple from correlation functions, the resulting
states that survive exactly match the gravity prediction for all
values of the 't Hooft coupling. 
The strict infinite $N$ limit, where the
sum of the number of boxes and antiboxes in the Young tableau 
has maximum value in the
conformal field theory partition function, is used.
The three point functions with scalars at tree level in
the undeformed bulk theory 
were computed  \cite{CY}.
In
particular, they have checked for spin-$3$ current and made
predictions for the three-point functions of spin $s \geq 4$, at
fixed 't Hooft coupling $\la=\frac{1}{2}$, in the $WA_{N-1}$ minimal model. 
In \cite{Ahn1111}, the three-point functions for spin-$4$ 
with scalars for all values of 't Hooft coupling
were found in the large $N$ 't Hooft limit of the $WA_{N-1}$ minimal model. 
The three-point functions with scalars in the deformed
bulk theory was found in \cite{AKP} and they were given by
scalar-scalar two-point functions. The result from conformal field
theory, along the line of \cite{GH}, agrees with the correlators from the bulk.   

There exist other types of minimal models in \cite{Ahn1106,GV}.
It is natural to ask what the three-point functions with two scalars in
these minimal models are, as described in \cite{Ahn1111} briefly \footnote{
For $WA_{N-1}$ coset model we have described in \cite{Ahn1111},
the coefficient functions in the 
coset primary spin-$4$ field were determined by the fact that
it should commute with the diagonal spin-$1$ current
and should transform as a primary field of dimension $4$ under the
coset stress energy tensor.
However, there remain two unknown and undetermined coefficient functions.
This implies that the above two requirements in the $WA_{N-1}$ coset
minimal model are not enough to determine all the coefficients in the
coset primary spin-$4$ field explicitly. 
As we described above, the field contents for the $W_N$ minimal model
for finite $N$ are given by the fields with spins $s=2, 3, 4, \cdots, N$.
By using the operator product expansion of coset primary spin-$3$ field with
itself and reading off the particular singular term $\frac{1}{(z-w)^2}$,
the above two unknown coefficients are fixed completely.}.

In this paper, 
we construct spin-$4$ primary Casimir operators in
$WB_{\frac{N-1}{2}}$ and $WD_{\frac{N}{2}}$ minimal models \cite{LF}.
Then we compute the three-point functions with two real scalars in the
large $N$ 't Hooft limit for all values of 't Hooft coupling.
The way in which the coset spin-$4$ currents in these minimal models are obtained is
rather different from the procedure that one uses for the $WA_{N-1}$
minimal model.
In \cite{Ahn1111}, we have constructed the coset primary spin-$4$ field by
considering the operator product expansion of coset primary spin-$3$ field
with itself (foonote $1$). These (spins $3$ and $4$) 
are two lowest higher spin currents in the $W_N$
minimal model. If one continues to compute the operator product
expansions between these lower spin currents, one obtains the higher
spin currents successively. 
In other words, one determines the spin-$3$ current using the above
two requirements and then fixes the spin-$4$ current. Then the
spin-$5$ current can be fixed from the operator product expansion
between the spin-$3$ current and the spin-$4$ current and so on. 

However, there exists an extra field of spin-$\frac{N}{2}$ for each case 
in the present minimal models. 
Depending on the $N$, this field is either fermionic or bosonic.
Then all the field contents of each minimal model are located at the
singular terms in the operator product expansion between this
spin-$\frac{N}{2}$ 
field and itself. Then the possible terms for primary spin-$4$ field can be
read off from the operator product expansion of this extra field with itself.
It is straightforward to apply the above two requirements for the
coset primary spin-$4$ field.
It turns out that all the coefficient functions are fixed except two
unknown coefficient functions. So far, the story looks similar to the
one for $WA_{N-1}$ minimal model.

Recall that the spin-$4$ field is the lowest higher spin field in each
minimal model. In other words, the lowest spin greater than $2$ is $4$.
In order to fix these two constants in terms of
$(N,k)$, one should compute the operator product expansion of primary spin-$4$
field with itself directly but it is almost impossible, by hand (or
other method), to compute these
quantities because the number of operator product expansions of
various spin-$4$ fields is $324$ for $WB_{\frac{N-1}{2}}$ minimal model 
and $452$ for $WD_{\frac{N}{2}}$ minimal model. 
Instead of doing this, what one can do, at the moment, is to 
resorts to the higher spin Lie algebra. From the eigenvalues for spin-$4$
zero mode in the higher spin Lie algebra, along the line of
\cite{GH,GGHR,AKP}, 
one can find the above two
undetermined coefficient functions (therefore all the coefficient
functions) 
in terms of $(N,k)$ in the large
$N$ 't Hooft limit. Of course, this is indirect approach but so far
there is no direct approach, in practice,  to fix the above two constants
\footnote{This feature is different from the behavior for the $W_N$ minimal
model where the eigenvalue equations for primary spin-$4$ zero mode 
satisfy the higher spin Lie algebra automatically.
By assuming the higher spin algebra for our primary spin-$4$ field
(i.e., the zero mode of primary spin-$4$ field should satisfy the wedge
subalgebra described in \cite{GH}), 
one obtains the complete structure for primary spin-$4$ field and the
three-point functions arise automatically.}.

In section 2, we review the Goddard-Kent-Olive (GKO) 
coset construction. 
We are interested in the specific minimal model characterized by the
coset central charge $\widetilde{c}$ that can be obtained from the
highest singular term in the 
operator product expansion of the above spin-$2$ coset Virasoro current.
 
In section 3, we consider the $WB_{\frac{N-1}{2}}$ minimal model. The
fundamental (generating) field content is given by the fermionic spin-$\frac{N}{2}$
field where $N$ is odd. All the field contents $s=2, 4, 6, \cdots,
(N-1)$
are obtained from the singular terms in the 
operator product expansion of this fundamental
field with itself. 
It turns out that there exist $18$ independent spin-$4$ fields which are
written in terms of two arbitrary coefficients.
From the eigenvalue equations of primary spin-$4$ zero
mode acting on the two primaries, 
we construct the three-point functions of
the spin-$4$ coset primary field with two real scalar fields in terms
of 't Hooft coupling constant under the assumption of higher spin Lie algebra. 

In section 4, we move on the $WD_{\frac{N}{2}}$ minimal model.
The generating field content is the bosonic spin-$\frac{N}{2}$ field
where $N$ is even. In this case, there exist $21$ independent spin-$4$
fields
where there are $3$ more spin-$4$ fields, compared to the previous case. 

In section 5, we summarize what we have obtained in this paper and we
make some comments on the future direction.

There are some (partial and incomplete) related works in \cite{GGR}-\cite{JJY}, 
along the line of \cite{GG}. 
       
\section{The GKO coset construction}

Let us consider the diagonal coset model \cite{GKO85,GKO86}
\bea
\frac{\widehat{SO}(N)_k \oplus \widehat{SO}(N)_1}{\widehat{SO}(N)_{k+1}}.
\label{coset}
\eea
The spin-$1$ fields $J^{ab}(z)$ and $K^{ab}(z)$, of level $k_1=1$ and $k_2=k$, 
generate the affine Lie algebra
$\widehat{SO}(N)_k \oplus \widehat{SO}(N)_1$.
The indices 
$a,b$ take the values $a,b = 1, 2, \cdots, N$
in the representation of finite dimensional Lie algebra $SO(N)$.
Due to the antisymmetric property of these fields ($J^{ab}(z) =-J^{ba}(z)$
and $K^{ab}(z)=-K^{ba}(z)$), the number of
independent fields is given by $\frac{1}{2}N(N-1)$ respectively.
The standard operator product expansions for these fields are given by
\bea
&& J^{ab} (z) J^{cd} (w)  =  -\frac{1}{(z-w)^2} k_1 (-\delta^{bc}
\delta^{ad}+
\delta^{ac} \delta^{bd}) 
\nonu \\ 
&& +  \frac{1}{(z-w)}
  \left[  \delta^{bc} J^{ad}(w) +\delta^{ad} J^{bc}(w)-\delta^{ac}
  J^{bd}(w)-
\delta^{bd} J^{ac}(w) \right] + \cdots,
\label{JJcoset}
\eea
and 
\bea
&& K^{ab} (z) K^{cd} (w)  =  -\frac{1}{(z-w)^2} k_2 (-\delta^{bc}
\delta^{ad}+
\delta^{ac} \delta^{bd}) 
\nonu \\ 
&& +  \frac{1}{(z-w)}
  \left[  \delta^{bc} K^{ad}(w) +\delta^{ad} K^{bc}(w)-\delta^{ac}
  K^{bd}(w)-
\delta^{bd} K^{ac}(w) \right] + \cdots.
\label{KKcoset}
\eea
When the pair of integers in terms of a single indices $A=(ab)$
and $B=(cd)$ is used, 
then the Kronecker delta is given by $\delta^{AB} =-\delta^{bc}
\delta^{ad}+
\delta^{ac} \delta^{bd}$. From this, $\delta^{AB}$ is symmetric under
the interchange of $A$ and $B$ (that is, $a \leftrightarrow c$ and $b
\leftrightarrow d$). 
However, under the change $ a \leftrightarrow b$ (or under the change
$c \leftrightarrow d$), 
it is antisymmetric. 
This is consistent with the operator product expansions (\ref{JJcoset}) and (\ref{KKcoset}).
Similarly, the structure constant 
with single index notation $f^{ABC}$ can be written as 
$f^{ABC}=f^{(ab)(cd)(ef)}=\delta^{ae} ( \delta^{bc}  \delta^{df}-\delta^{bd} \delta^{cf} )  
+\delta^{be} ( \delta^{ad} \delta^{cf} -\delta^{ac} \delta^{df} )$
\footnote{ \label{footnote2}
Then the operator product expansion (\ref{JJcoset}) can be written as 
$J^A(z) J^B(w) =-\frac{1}{(z-w)^2} k_1 \delta^{AB} +\frac{1}{(z-w)} f^{ABC}
J^C(w) +\cdots$ and similarly, the operator product expansion
(\ref{KKcoset}) can be written as 
$K^A(z) K^B(w) =-\frac{1}{(z-w)^2} k_2 \delta^{AB} +\frac{1}{(z-w)} f^{ABC}
K^C(w) +\cdots$.
Here, the $SO(N)$ generator can be realized as $N \times N$ matrix with
the $(i,j)$ component  
$T^A_{ij} = T^{(ab)}_{ij} = i (\delta^a_i \delta^b_j - \delta^a_j
\delta^b_i )$. One obtains $\mbox{Tr} (T^A T^B)=2
\delta^{AB}$ by taking the trace for the product of two generators,
$T^{A}_{ij} T^{A}_{kl}=-\delta_{ik} \delta_{jl}+\delta_{il}
\delta_{jk}$ by summing over the indices $a, b =1, 2, \cdots, N$ (dividing by
$2$) 
where 
the generators satisfy the usual commutator relation $[T^A, T^B] = i f^{ABC} T^C$.}.

The spin-$1$ field of the diagonal affine Lie subalgebra $\widehat{SO}(N)_{k+1}$
in the coset model (\ref{coset}) is the sum of $J^{ab}(z)$ and $K^{ab}(z)$
\bea
J'^{ab} (z) = J^{ab}(z) +K^{ab}(z).
\label{diag}
\eea
The operator product expansion of (\ref{diag}) can be obtained from the defining
equations (\ref{JJcoset}) and (\ref{KKcoset}) and the fact that there
are no singular terms in the operator product expansion $J^{ab}(z) K^{cd}(w)$:
\bea
J'^{ab} (z) J'^{cd} (w) & = & 
-\frac{1}{(z-w)^2} (k_1+k_2) (-\delta^{bc}
\delta^{ad}+
\delta^{ac} \delta^{bd}) 
\nonu \\ 
& + & \frac{1}{(z-w)}
  \left[  \delta^{bc} J'^{ad}(w) +\delta^{ad} J'^{bc}(w)-\delta^{ac}
  J'^{bd}(w)-
\delta^{bd} J'^{ac}(w) \right] + \cdots.
\label{opej'j'}
\eea
The level of the field $J'^{ab}(z)$ is the sum of $k_1$ and $k_2$, $k'=k_1+k_2=1+k$.
In a sence, the coset model (\ref{coset}) can be viewed as 
perturbations of the $k \rightarrow \infty$ model.

The above GKO construction looks very similar to those for the coset model
$\frac{\widehat{SU}(N)_k \oplus
  \widehat{SU}(N)_1}{\widehat{SU}(N)_{k+1}}$.
The only difference appears in both the dual Coxeter number and the dimension
of group if we use a single index notation.  
In next sections, we would like to construct the two lowest higher spin
generators (extending the spin-$2$ coset construction in the Appendix
$A$ to the 
higher spin currents) from the spin-$1$ fields $J^{ab}(z)$ and $K^{ab}(z)$.  

\section{The fourth-order Casimir operator of
  $B_{\frac{N-1}{2}}=SO(N)$ where $N$ is odd}

In this section, we construct the spin-$4$ primary field, after
that we take the large $N$ limit, and compute the three-point functions
with scalars.

\subsection{ Primary spin-4 current}

The $WB_{\frac{N-1}{2}}$ algebra by Fateev and Lukyanov 
\cite{LF} is generated by the fields of spins 
\bea
2, 4, \cdots, (N-1), \frac{N}{2}, \qquad N: \mbox{odd}.
\label{series}
\eea
The orders of the independent Casimir operators for the 
non-simply-laced simple Lie algebra $B_{\frac{N-1}{2}}$ 
are given by $2, 4, \cdots, (N-1)$. 
The spin contents of $WB_{\frac{N-1}{2}}$ algebra are related to 
the exponent of the Lie superalgebra $B(0,\frac{N-1}{2})=OSp(1,N-1)$.
The operator product expansion of fermionic spin-$\frac{N}{2}$ field
with itself provides
the structure of the remaining bosonic fields of spin $2, 4, \cdots, (N-1)$:
\bea
\widetilde{U}_{WB}(z) \widetilde{U}_{WB}(w) & = & \frac{1}{(z-w)^N} \frac{2\widetilde{c}}{N}+
\frac{1}{(z-w)^{N-2}} 2 \widetilde{T}(w) +
\frac{1}{(z-w)^{N-3}} \pa \widetilde{T}(w)
\nonu \\
&+ & \frac{1}{(z-w)^{N-4}}  \left[ \widetilde{T} \widetilde{T}(w),
  \pa^2 \widetilde{T}(w), \widetilde{V}_{WB}(w) \right]
+ {\cal O}((z-w)^{-N+5}).
\label{uu}
\eea
The bosonic currents can be read off from (\ref{uu}).
The coset central charge $\widetilde{c}$ is given by (A.4).
The spin-$2$ field appears in the singular term
$\frac{1}{(z-w)^{N-2}}$ and its descendant spin-$3$ field is located
at the next singular term $\frac{1}{(z-w)^{N-3}}$.
The terms in $\frac{1}{(z-w)^{N-4}}$ of (\ref{uu}) have 
spin-$4$ fields. We would like to find the spin-$4$ primary field 
$\widetilde{V}_{WB}(z)$ explicitly. One should also consider the 
spin-$4$ fields $\widetilde{T} \widetilde{T}(z)$ and 
$\pa^2 \widetilde{T}(z)$ 
coming from the stress energy tensor $\widetilde{T}(z)$.
In principle, the higher spin fields of spin $s$ greater than $4$
arise in the lower singular terms in (\ref{uu}) but its exact
structure is not known explicitly so far. The highest spin 
field with spin-$(N-1)$ in (\ref{series}) 
appears in the $\frac{1}{(z-w)}$ term which is the
lowest singular term. Since there is no spin-$1$ field in this minimal
model, there is no $\frac{1}{(z-w)^{N-1}}$ term in the operator
product expansion (\ref{uu}). 
For $N=3$, the $WB_1$ algebra coincides with the ${\cal N}=1$ super
Virasoro algebra.

It is ready to construct the above spin-$\frac{N}{2}$ field in terms
of spin-$1$ fields $J^{ab}(z)$ and $K^{ab}(z)$.
One defines the spin-$1$ field as composite of the $N$-free fermions \cite{Watts}
\bea
J^{ab}(z) = \psi^a \psi^b(z).
\label{twofermion}
\eea
The operator product expansion of fermions of spin $s=\frac{1}{2}$ is 
\bea
\psi^a(z) \psi^b(w) =\frac{1}{(z-w)} \delta^{ab} +\cdots.
\label{opepsipsi}
\eea
The fermion fields anticommute and have the mode expansion with
the Neveu-Schwarz sector or the Ramond sector.
It is easy to check the operator product expansion 
(\ref{JJcoset}) is satisfied with level $k_1=1$ by using
(\ref{twofermion}) 
and (\ref{opepsipsi}).
One also checks that this  fermion is primary field of
spin-$\frac{1}{2}$ 
under the stress
energy tensor $T_{(1)}(z) = -\frac{1}{2} \psi^a \pa \psi^a(z)$ in (A.2).
According to the observation of Watts \cite{Watts}, the spin-$\frac{N}{2}$ field
$\widetilde{U}(z)$ consists of $\frac{N+1}{2}$ independent terms with
arbitrary coefficient functions $A$'s which depend on both $N$ and $k$
(the explicit expressions for these coefficients are given in \cite{Watts})
\bea
\widetilde{U}_{WB}(z) & = &
\ep^{a_1 a_2 \cdots a_N} \left[ A_0(N,k)  \psi^{a_1} J^{a_2 a_3} \cdots J^{a_{N-1}
a_N}(z) + \cdots  \right. \nonu \\
&+ &  A_{\frac{N-i-1}{2}}(N,k)  \psi^{a_1} J^{a_2 a_3} \cdots J^{a_i
  \, a_{i+1}} K^{a_{i+2} \, a_{i+3}}
\cdots K^{a_{N-1} a_N}(z) \nonu \\
&+ &  \left.  \cdots + 
A_{\frac{N-1}{2}}(N,k) \psi^{a_1} K^{a_2 a_3} 
\cdots K^{a_{N-1} a_N}(z) \right].
\label{fermionU}
\eea
This is a singlet under the underlying $SO(N)$ subalgebra of
$\widehat{SO}(N)$.
The epsilon tensor of $N$ indices  is $SO(N)$ group invariant.
Now we substitute (\ref{fermionU}) into the operator product expansion 
(\ref{uu}) and look for $\frac{1}{(z-w)^{N-4}}$ terms.  
Using the operator product expansions (\ref{JJcoset}) and (\ref{KKcoset}),
the four indices in the left hand side will distribute to 
either Kronecker delta $\delta^{ef}$ or spin-$1$ fields $J^{ef}(z)$ or
$K^{ef}(z)$ in the right hand side.
At first, one sees the lowest singular term $\frac{1}{(z-w)}$ in the
operator product expansion $\widetilde{U}_{WB}(z) \widetilde{U}_{WB}(w)$.
The higher singular terms ($\frac{1}{(z-w)^n}$ where $n =2, 3, 4,
\cdots, N$) can be obtained from the spin-$(N-1)$ field 
located at $\frac{1}{(z-w)}$ term 
by contracting the remaining indices between the fields in the normal
ordered product.
For example, the operator product expansion between the first term of 
(\ref{fermionU}) with itself will lead to 
$\ep^{a_1 a_2 \cdots a_N} \ep^{a_1 b_2 \cdots b_N}  (J^{a_2 a_3} \cdots J^{a_{N-1}
a_N})( J^{b_2 b_3} \cdots J^{b_{N-1}
b_N})(w)$ after using the operator product expansion between
the field $\psi^{a_1}(z)$
and the field $\psi^{b_1}(w)$.
Further contractions between the remaining expression will give rise to 
the lower spin field of spin $(N-2)$ by removing one current or spin
$(N-3)$ by removing two currents.  
Then the Kronecker delta's make a contraction between two $SO(N)$
epsilon tensors and the order of the original spin-$1$ fields, $(N-1)$, is reduced
to $(N-2), (N-3),
(N-4) \cdots, 4, 3, 2, 0$ depending on the
location of singular terms where the descendant fields are added. 
The fields of spins $3, 5, \cdots, (N-2)$ will correspond to the
descendant fields of bosonic fields of spins $2, 4, \cdots, (N-1)$ 
of $WB_{\frac{N-1}{2}}$ minimal model. 

How does one determine the nontrivial spin-$4$ field which has the lowest higher
spin greater than $2$ in the $WB_{\frac{N-1}{2}}$ minimal model?
It is easy to see, after completing the procedure in previous
paragraph, that the spin-$2$ fields coming from
$\widetilde{U}_{WB}(z)$
are given by $J^{cd} J^{ef}(z), J^{cd} K^{ef}(z)$ and $K^{cd} K^{ef}(z)$
with epsilon tensor $ \ep^{a_1 a_2 \cdots a_{N-4} c d e f} $
and similarly those from $\widetilde{U}_{WB}(w)$ are 
$J^{gh} J^{ij}(w), J^{gh} K^{ij}(w)$ and $K^{gh} K^{ij}(w)$ with epsilon tensor
$\ep^{b_1 b_2 \cdots b_{N-4} g h i j}$.
The normal ordered products of these fields with appropriate
contracted two epsilon tensors arise in the 
singular term $\frac{1}{(z-w)^{N-4}}$ in (\ref{uu}).
It turns out that the spin-$4$ fields have the following structure
\bea
&& \ep^{a_1 a_2 \cdots a_{N-4} c d e f} 
\ep^{a_1 a_2 \cdots a_{N-4} g h i j} (J^{cd} J^{ef}) (J^{gh} J^{ij})(z) \sim 
\delta^c_{[g} \delta^d_h \delta^e_i \delta^f_{j]} (J^{cd} J^{ef}) (J^{gh} J^{ij})(z),
\label{Spinspin4}
\\
&& \ep^{a_1 a_2 \cdots a_{N-4} c d e f} 
\ep^{a_1 a_2 \cdots a_{N-4} g h i j} (J^{cd} J^{ef}) (J^{gh} K^{ij})(z) \sim 
\delta^c_{[g} \delta^d_h \delta^e_i \delta^f_{j]}  (J^{cd} J^{ef}) (J^{gh} K^{ij})(z),
\nonu \\
&& \ep^{a_1 a_2 \cdots a_{N-4} c d e f} 
\ep^{a_1 a_2 \cdots a_{N-4} g h i j} (J^{cd} J^{ef}) (K^{gh} K^{ij})(z) \sim
\delta^c_{[g} \delta^d_h \delta^e_i \delta^f_{j]} (J^{cd} J^{ef}) (K^{gh} K^{ij})(z),
\nonu \\
&& \ep^{a_1 a_2 \cdots a_{N-4} c d e f} 
\ep^{a_1 a_2 \cdots a_{N-4} g h i j} (J^{cd} K^{ef}) (J^{gh} K^{ij})(z) \sim
\delta^c_{[g} \delta^d_h \delta^e_i \delta^f_{j]} (J^{cd} K^{ef}) (J^{gh} K^{ij})(z),
\nonu \\
&& \ep^{a_1 a_2 \cdots a_{N-4} c d e f} 
\ep^{a_1 a_2 \cdots a_{N-4} g h i j} (J^{cd} K^{ef}) (K^{gh} K^{ij})(z) \sim
\delta^c_{[g} \delta^d_h \delta^e_i \delta^f_{j]} (J^{cd} K^{ef}) (K^{gh} K^{ij})(z),
\nonu \\
&& \ep^{a_1 a_2 \cdots a_{N-4} c d e f} 
\ep^{a_1 a_2 \cdots a_{N-4} g h i j} (K^{cd} K^{ef}) (K^{gh} K^{ij})(z) \sim
\delta^c_{[g} \delta^d_h \delta^e_i \delta^f_{j]} (K^{cd} K^{ef}) (K^{gh} K^{ij})(z),
\nonu
\eea
where the contractions between two epsilon tensors are proportional to
the Kronecker deltas. 
The right hand side of (\ref{Spinspin4}) can be further simplified by
removing the indices $g,h,i$ and $j$.
Then one writes down the possible various spin-$4$ fields, by
simplifying the right hand side of (\ref{Spinspin4}), as follows: 
\bea
&& c_1 J^{cd} J^{ef} J^{cd} J^{ef}(z) + c_2 J^{cd} J^{ef} J^{cd} K^{ef}(z)+
c_3 J^{cd} J^{ef} K^{cd} K^{ef}(z)+ c_4 J^{cd} K^{ef} K^{cd} K^{ef}(z) 
\nonu \\
&& + c_5 K^{cd} K^{ef} K^{cd} K^{ef}(z) 
+
c_6 J^{cd} J^{ef} J^{ce} J^{df}(z) + c_7 J^{cd} J^{ef} J^{ce} K^{df}(z)+
c_8 J^{cd} J^{ef} K^{ce} K^{df}(z)  \nonu \\
&& + c_9 J^{cd} K^{ef} K^{ce} K^{df}(z) +
c_{10} K^{cd} K^{ef} K^{ce} K^{df}(z) +
c_{11} J^{cd} J^{cd} J^{ef} J^{ef}(z) + c_{12} J^{cd} J^{cd} J^{ef} K^{ef}(z)
\nonu \\
&& + c_{13} J^{cd} J^{cd} K^{ef} K^{ef}(z) + 
 c_{14} J^{cd} K^{cd} K^{ef} K^{ef}(z) +
c_{15} K^{cd} K^{cd} K^{ef} K^{ef}(z) +
c_{16} J^{cd} J^{ce} J^{ef} J^{df}(z)   \nonu \\
&&  + c_{17} J^{cd} J^{ce} J^{ef} K^{df}(z) +
c_{18} J^{cd} J^{ce} K^{ef} K^{df}(z) + 
c_{19} J^{cd} K^{ce} K^{ef} K^{df}(z) +
c_{20} K^{cd} K^{ce} K^{ef} K^{df}(z) 
\nonu \\
&& + c_{21} J^{cd} J^{ce} J^{df} J^{ef}(z) + c_{22} J^{cd} J^{ce} J^{df}
K^{ef}(z) +
c_{23} J^{cd} J^{ce} K^{df} K^{ef}(z) + 
c_{24} J^{cd} K^{ce} K^{df} K^{ef}(z) 
\nonu \\
&& + c_{25} K^{cd} K^{ce} K^{df} K^{ef}(z).
\label{spin4quartic}
\eea
The spin-$4$ fields are quartic in the currents with appropriate index structure.
Of course, one should expect that there 
exist some derivative terms between the spin-$1$ fields from the
normal ordered products (\ref{Spinspin4}) to fully normal ordered
products
(\ref{spin4quartic}). 
Compared to the minimal model based on $SU(N)$ group where the $d$
symbols of different ranks are contracted with the currents, 
the symmetric $SO(N)$ invariant tensor of rank $2$,
Kronecker delta, plays an important role.   

On the other hand, one can think of the following derivatives
\bea
&&
d_1 \pa J^{ab} \pa J^{ab}(z)  
+ d_2 \pa^2 J^{ab}  J^{ab}(z) 
+ d_3 \pa K^{ab} \pa K^{ab}(z)  
+ d_4 \pa^2 K^{ab}  K^{ab}(z) 
\nonu \\
&& + d_5
\pa^2 J^{ab}  K^{ab}(z) 
+ d_6 \pa J^{ab} \pa K^{ab}(z) 
+
d_7 J^{ab} \pa^2 K^{ab}(z) 
+ d_8 
J^{ab} \pa J^{ac} K^{bc}(z) 
\nonu \\
&& + d_9
J^{ab} K^{ac} \pa K^{bc}(z),
\label{spin4derivative}
\eea
where some of these come from the derivative field of stress energy
tensor
$\pa^2 \widetilde{T}(z)$.

Therefore, the spin-$4$ candidate given by (\ref{spin4quartic}) and 
(\ref{spin4derivative})
can be further simplified and summarized by the following
$21(=25+9-13)$ (the 13 terms can be written as other terms from the
footnote 4)
independent terms, via the detailed analysis in the Appendix $B$,
\bea
&& 
\widetilde{V}(z)= c_3 J^{cd} J^{ef} K^{cd} K^{ef}(z)+
c_8 J^{cd} J^{ef} K^{ce} K^{df}(z)+ c_9 J^{cd} K^{ef} K^{ce} K^{df}(z)
\nonu \\ 
&& +
c_{10} K^{cd} K^{ef} K^{ce} K^{df}(z)
 +
c_{11} J^{cd} J^{cd} J^{ef} J^{ef}(z) + c_{12} J^{cd} J^{cd} J^{ef} K^{ef}(z)
 + c_{13} J^{cd} J^{cd} K^{ef} K^{ef}(z) \nonu \\
&& + 
 c_{14} J^{cd} K^{cd} K^{ef} K^{ef}(z) 
 +
c_{15} K^{cd} K^{cd} K^{ef} K^{ef}(z) +
c_{18} J^{cd} J^{ce} K^{ef} K^{df}(z) + c_{21} J^{cd} J^{ce} J^{df}
J^{ef}(z) \nonu \\
&& 
+ c_{22} J^{cd} J^{ce} J^{df}
K^{ef}(z) 
+ d_1 \pa J^{ab} \pa J^{ab}(z)  
+ d_2 \pa^2 J^{ab}  J^{ab}(z) 
+ d_3 \pa K^{ab} \pa K^{ab}(z)  
+ d_4 \pa^2 K^{ab}  K^{ab}(z) 
\nonu \\
&& + d_5
\pa^2 J^{ab}  K^{ab}(z) 
+ d_6 \pa J^{ab} \pa K^{ab}(z) 
+
d_7  J^{ab} \pa^2 K^{ab}(z) 
+ d_8 
J^{ab} \pa J^{ac} K^{bc}(z) 
\nonu \\
&& + d_9
J^{ab} K^{ac} \pa K^{bc}(z).
\label{Vtildeinter}
\eea
Compared to the $WB_2$ minimal model (i.e., $N=5$) \cite{Ahn1992} (See
also \cite{Ahn1991}),
there exist three extra terms: $c_8$-term, $c_{21}$-term and $c_{22}$-term.
We will see that these extra terms can be absorbed into the other
independent terms for $WB_{\frac{N-1}{2}}$ minimal model by using 
$N$-free fermion description.
We also use these $21$ independent terms for the $WD_{\frac{N}{2}}$
minimal model.

It is ready to determine the coeffcient functions in (\ref{Vtildeinter}).
At first, the primary spin-$4$ field should commute with the diagonal
spin-$1$ field as follows \cite{BBSS1}:
\bea
J'^{ab}(z) \widetilde{V}(w) = \mbox{regular}.
\label{regular}
\eea
In other words, there are no singular terms ($\frac{1}{(z-w)^n}$ terms
where $n=5,4,3,2,1$) in the operator product
expansion (\ref{regular}).
Secondly, the coset spin-$4$ primary field should transform
as dimension 4 under the stress energy tensor (A.1) 
as follows \cite{BBSS1}:
\bea
\widetilde{T}(z) \widetilde{V}(w) = \frac{1}{(z-w)^2} 4
\widetilde{V}(w) +\frac{1}{(z-w)} \pa \widetilde{V}(w)
+\cdots.
\label{tprimary}
\eea
That is, there should be no singular terms 
($\frac{1}{(z-w)^n}$ terms
where $n=6,5,4,3$) in the operator product
expansion (\ref{tprimary}).
Sometimes, it is convenient to introduce
the stress energy tensor in the affine Lie algebra  
$\widehat{SO}(N)_k \oplus \widehat{SO}(N)_1$:
\bea
T_{(1)}(z) + T_{(2)}(z) \equiv \hat{T}(z).
\label{That}
\eea
The equation (\ref{regular}) implies that 
there are no singular terms in the operator product expansion of
$T'(z) \widetilde{V}(w)$ because $T'(z)$ is quadratic in $J'^{ab}(z)$
from (A.2). Therefore, it is equivalent to compute the
operator product expansion of $\hat{T}(z) \widetilde{V}(w)$.  
In the Appendix
 $C$, we describe the operator product expansions 
(\ref{regular}) where we consider the spin-$4$ field in
(\ref{spin4quartic})
and (\ref{spin4derivative}).
The reason for why we do take these rather than 
(\ref{Vtildeinter}) is that sometimes 
we want to express the spin-$4$ field which is quartic in the currents
without any derivative terms. 
In the Appendix $D$, we compute the operator product expansion 
$\hat{T}(z) \widetilde{V}(w)$ which should be 
equal to the equation (\ref{tprimary}) under the condition (\ref{regular}). 
We will describe some details in the Appendix $B$. 

Therefore, we take the final correct spin-$4$ field as follows:
\bea
&& 
\widetilde{V}_{WB}(z)= c_3 J^{cd} J^{ef} K^{cd} K^{ef}(z)+
 c_9 J^{cd} K^{ef} K^{ce} K^{df}(z) +
c_{10} K^{cd} K^{ef} K^{ce} K^{df}(z) \nonu \\
&& +
c_{11} J^{cd} J^{cd} J^{ef} J^{ef}(z) + c_{12} J^{cd} J^{cd} J^{ef} K^{ef}(z)
 + c_{13} J^{cd} J^{cd} K^{ef} K^{ef}(z) + 
 c_{14} J^{cd} K^{cd} K^{ef} K^{ef}(z) \nonu \\
&& +
c_{15} K^{cd} K^{cd} K^{ef} K^{ef}(z) +
c_{18} J^{cd} J^{ce} K^{ef} K^{df}(z)  
+ d_1 \pa J^{ab} \pa J^{ab}(z)  
+ d_2 \pa^2 J^{ab}  J^{ab}(z) \nonu \\
&& + d_3 \pa K^{ab} \pa K^{ab}(z)  
+ d_4 \pa^2 K^{ab}  K^{ab}(z) 
+ d_5
\pa^2 J^{ab}  K^{ab}(z) 
+ d_6 \pa J^{ab} \pa K^{ab}(z) 
+
d_7  J^{ab} \pa^2 K^{ab}(z) 
\nonu \\
&& + d_8 
J^{ab} \pa J^{ac} K^{bc}(z) + d_9
J^{ab} K^{ac} \pa K^{bc}(z).
\label{Vtilde}
\eea
For $N=5$, the field contents of (\ref{Vtilde}) are exactly same as 
the ones in \cite{Ahn1992}.
This is one of the reasons why we take the particular combination 
for the various spin-$4$ fields as in (\ref{Vtilde}).  
In Appendix $E$, the requirements (\ref{regular}) and (\ref{tprimary})
are imposed and the coefficient functions appearing the spin-$4$ field
in (\ref{Vtilde}), in terms of finite $(N,k)$, are determined. 
However, the coefficient functions $c_9$ and $d_8$ are not fixed.
All the coefficient functions are written in terms of these 
two coefficient functions.
This common feature also occurs in the $W_N$ minimal model if one
does not consider the operator product expansion between the primary
spin-$3$
fields. According to the field contents of (\ref{series}), 
there are no lower spin fields of spin less than $4$, contrary to the
$WA_{N-1}$ minimal model. In order to fix these unknown coefficient
functions, one should compute the operator product expansion
of $\widetilde{V}_{WB}(z) \widetilde{V}_{WB}(w)$ explicitly.  
In the Appendix $F$, we present the field contents for the $WB_2$
algebra corresponding to $N=5$ case for convenience.

The primary spin-$4$ current which is fourth order Casimir operator of
$SO(N)$ where $N$ is odd is given by 
(\ref{Vtilde}) with the coefficient functions in (E.7).
In next subsection, we describe this primary spin-$4$ current in the
large $N$ limit and find three-point functions with scalars. 

\subsection{ Primary spin-$4$ current in the large $N$ 't Hooft limit and
  three-point functions with two scalars }

The large $N$ 't Hooft limit is described as \cite{GG}
\bea
N, k \rightarrow \infty, \qquad \lambda \equiv \frac{N}{N+k}
\qquad \mbox{fixed}.
\label{limit}
\eea

One should find the spin $4$ zero mode on the vector representation.
The spin-$4$ field is given by (\ref{Vtilde}).
Let us first consider the quartic terms.
From the matrix representation in the footnote \ref{footnote2}, one has 
\bea
\mbox{Tr}(T^{cd} \, T^{ef} \, T^{cd} \, T^{ef}) & = &
 i (\delta^c_i \delta^d_j - \delta^c_j
\delta^d_i )
 i (\delta^e_j \delta^f_k - \delta^e_k
\delta^f_j )
 i (\delta^c_k \delta^d_l - \delta^c_l
\delta^d_k )
 i (\delta^e_l \delta^f_i - \delta^e_i
\delta^f_l ) \nonu \\
&= &
4N(N-1) \rightarrow 4N^2. 
\label{above}
\eea
Here we take the large $N$ limit (\ref{limit}).
In order to obtain the eigenvalue, one should divide this (\ref{above}) by $N$.
Then, the zero mode (relevant to the $c_1$-$c_5$ terms) 
acting on the vector representation implies that 
\footnote{Similarly, 
from the identity
$
\mbox{Tr} (T^{cd} \, T^{ef} \, T^{ce} \, T^{df}) =
N^2(N-1) \rightarrow N^3$ that can be obtained from the matrix
representation for the generator, 
one obtains 
\bea
J_0^{cd} J_0^{ef} J_0^{ce} J_0^{df} |v> = N^2 |v>.
\label{6-10dep}
\eea
which are relevant to the $c_6$-$c_{10}$ terms.
It is straightforward to compute
$
\mbox{Tr}(T^{cd} \, T^{cd} \, T^{ef} \, T^{ef}) =
4N(N-1)^2 \rightarrow 4N^3$ which can be checked from the matrix representation, 
and the corresponding eigenvalue equation (relevant to
$c_{11}$-$c_{15}$ terms) leads to
\bea
J_0^{cd} J_0^{cd} J_0^{ef} J_0^{ef} |v> = 4N^2 |v>.
\label{11-15dep}
\eea
Furthermore, 
from the relation
$
\mbox{Tr}(T^{cd} \, T^{ce} \, T^{ef} \, T^{df}) =
N(N-1)(N^2-3N+4) \rightarrow N^4$, 
one has 
\bea
J_0^{cd} J_0^{ce} J_0^{ef} J_0^{df} |v> = N^3 |v>,
\label{16-20dep}
\eea
which are relevant to $c_{16}-c_{20}$ terms.
The last one one should have is 
$
\mbox{Tr}(T^{cd} \, T^{ce} \, T^{df} \, T^{ef}) =
N^2(N-1) \rightarrow N^3$, 
and the eigenvalue equation (corresponding to $c_{21}$-$c_{25}$ terms) is given by
\bea
J_0^{cd} J_0^{ce} J_0^{df} J_0^{ef} |v> = N^2 |v>.
\label{21-25dep}
\eea
Therefore, the power of $N$ in (\ref{16-20dep}) is higher than the
ones in (\ref{1-5dep}), (\ref{6-10dep}), (\ref{11-15dep}) or (\ref{21-25dep}). 
The $N^3$ behavior of (\ref{16-20dep}) is the same as the one in
$WA_{N-1}$ minimal model \cite{Ahn1111}.

Let us consider the quadratic and cubic terms in (\ref{Vtilde}). 
One needs to have 
\bea
\mbox{Tr}(T^{cd} \, T^{cd}) = -2N(N-1) \rightarrow -2N^2,
\label{ttrel}
\eea
which is relevant to the $d_1$-$d_9$ terms  
and dividing this (\ref{ttrel}) by $N$, one obtains
spin-$2$ zero mode on the vector representation in the large $N$ 't
Hooft limit
\bea
J_0^{cd} J_0^{cd} |v> = -2N |v>.
\label{quadraticdep}
\eea
From (\ref{quadraticdep}), one can obtain the spin-$4$ zero mode with
two derivatives.}
\bea
J_0^{cd} J_0^{ef} J_0^{cd} J_0^{ef} |v> = 4N |v>.
\label{1-5dep}
\eea

Combining the results in the Appendix $E$, 
the leading contribution $N^2$ from $d_8$ factor comes from the coeffcient
functions,
$c_{18}$, $d_1$, $d_2$, $d_5$, $d_6$, $d_7$, $d_8$ and $d_9$
\bea
&& N^3 c_{18} - 2N d_1 - 4N d_2 
+4N d_5 + 2N d_6 + 4N d_7  
 -N^2 d_{8}+
N^2 d_9   \nonu \\
&& \rightarrow 
-\left[ \frac{ N^2 \left(-12-16 \lambda -99 \lambda ^2+85 \lambda
    ^3\right)}{10 (-2+\lambda ) \lambda  (-6+5 \lambda )} \right] d_8.
\label{equation1-1}
\eea
The leading contribution $N^3$ from $c_9$ factor comes from 
the coefficient functions $d_7$ and $d_9$.
\bea
4N d_7  +
N^2 d_9  \rightarrow 
-
\left[ \frac{14
   N^3 (-1+\lambda )}{-6+5 \lambda } \right] c_9.
\label{equation2}
\eea
Finally, by substituting the coefficient functions in the large $N$
limit
into (\ref{spin4quartic}) and (\ref{spin4derivative}) and evaluating
the correct eigenvalues, one arrives at the final contributions 
acting on the representation $( v ; 0) \otimes ( v ; 0)$, where 
$J_0^{ab} +K_0^{ab} =0$,
by combining (\ref{equation1-1}) and (\ref{equation2}), 
\bea
\widetilde{V}_0 | {\cal{O}}_{+} > & = & 
-\frac{N^2}{(-6+5 \lambda )}\left[ 
\frac{d_{8} \left(-12-16 \lambda -99 \lambda ^2+85 \lambda
    ^3\right)}{10 (-2+\lambda ) \lambda  } \right.
\nonu \\
& + & \left. 14
  c_{9} N (-1+\lambda )\right] 
| {\cal{O}}_{+} >,
\label{eigen1}
\eea
where
$ {\cal{O}}_{+} \equiv 
( v ; 0) \otimes ( v ; 0)$ and this is equivalent to $ (2,
1^{\frac{N-3}{2}}|1^{\frac{N-1}{2}}) 
\otimes (2,
1^{\frac{N-3}{2}}|1^{\frac{N-1}{2}}) $ in the
convention of \cite{Ahn1106}.

Next let us consider the zero mode eigenvalue  
acting on $ |{\cal{O}}_{-} > \equiv 
| ( 0; v ) \otimes ( 0; v ) >$.
For the primary $( 0; v ) \otimes ( 0; v )$, the field $K_0^{ab}$
vanishes.
Then there exist nonzero contributions from $c_{11}$-, $d_1$- and $d_2$-terms.
The $c_{11}$-term has $N^2 \times \frac{1}{N}=N$ dependence.
The $d_1$- and $d_2$-terms have $N \times N=N^2$ dependence.
So one arrives at 
\bea
-2N d_1 -4N d_2  =- \left[ \frac{ N^2 (-1+\lambda )}{10 \lambda }
  \right] d_8.
\nonu
\eea
In other words, one obtains
\bea
\widetilde{V}_0 | {\cal{O}}_{-} > & = &  -N^2 \left[
\frac{  (-1+\lambda )}{10 \lambda } \right] d_8
 | {\cal{O}}_{-} >, 
\label{eigen2}
\eea
where
$ {\cal{O}}_{-} \equiv 
( 0; v ) \otimes ( 0; v )$ which is equal to  $ (1^{\frac{N-1}{2}}|2,
1^{\frac{N-3}{2}}) 
\otimes (1^{\frac{N-1}{2}}|2,
1^{\frac{N-3}{2}}) $ in the
convention of \cite{Ahn1106}.
The vector representation of $SO(N)$ is self-conjugate and there is no
separate conjugate representation, contrary to the fundamental
representation of $SU(N)$.
  
For the choice of
\bea
c_{9}(N,\la) = -\frac{15 (-2+\lambda ) \lambda ^2 (-5+3 \lambda )}{7 N^3
  (-1+\lambda )}, \qquad
d_{8}(N,\la) =  \frac{10 (-3+\lambda ) (-2+\lambda )
  \lambda }{N^2},
\label{twoconstant}
\eea
one has the following eigenvalue equations, from (\ref{eigen1}) and (\ref{eigen2}),
\bea
\widetilde{V}_0 | {\cal{O}}_{+} > & = &   
(1+\lambda)(2+\lambda)(3+\lambda)
| {\cal{O}}_{+} >, \qquad {\cal{O}}_{+} \equiv 
( v ; 0) \otimes ( v ; 0), \nonu \\
\widetilde{V}_0 | {\cal{O}}_{-} > & = &   
(1-\lambda)(2-\lambda)(3-\lambda)
| {\cal{O}}_{-} >, \qquad  {\cal{O}}_{-} \equiv 
( 0; v ) \otimes ( 0; v ).
\label{EEigeneigen}
\eea
Recall that the zero mode eigenvalues 
for arbitrary spin $s$ in the boundary theory are found in
\cite{AKP}.
If one puts $s=4$, then they are exactly the same as
(\ref{EEigeneigen})
up to unfixed $\la$-independent normalization which depends on the
spin $s$ explicitly.
One expects that if one computes the operator product expansion 
$\widetilde{V}_{WB}(z) \widetilde{V}_{WB}(w)$, the singular terms should behave
as $\widetilde{V}_{WB}(z) \widetilde{V}_{WB}(w)=
\frac{1}{(z-w)^8} \frac{\widetilde{c}}{4} +\frac{1}{(z-w)^6} 2
\widetilde{T}(w) +\frac{1}{(z-w)^5} \pa \widetilde{T}(w) +{\cal
  O}(\frac{1}{(z-w)^{4}})$ where the stress energy tensor is given by
(A.1)
and (A.2). Then the undetermined two coefficient
functions $c_9(N,k)$ and $d_8(N,k)$ occur in this operator product
expansion. Only after this computation which will be very complicated
(i.e., $18 \times 18= 324$ operator product expansions one should compute) is done, 
they are fixed completely. Otherwise, one does not know what they are.
They should take the form
(\ref{twoconstant}) as one takes the large $N$ limit. 

The three-point functions with two real scalars, from
(\ref{EEigeneigen}),  
is summarized as
\bea
<{\cal{O}}_{+} {\cal{O}}_{+} \widetilde{V}> & = & 
(1+\lambda)(2+\lambda)(3+\lambda), \nonu \\
<{\cal{O}}_{-} {\cal{O}}_{-} \widetilde{V}> & = & 
(1-\lambda)(2-\lambda)(3-\lambda).
\label{three}
\eea
It would be interesting to find the three-point functions in the
deformed $AdS_3$ bulk theory for all values of 't Hooft coupling
constant
and to compare to the three-point
functions (\ref{three}) in the $WB_{\frac{N-1}{2}}$ coset conformal
field theory in the large $N$ limit. See, for example, \cite{AKP}.

We present the final spin-$4$ primary field with $(N,k)$ dependent  
coefficient functions  in (E.14). In the large $N$ limit, 
this becomes further simple expression as follows:  
\bea
\widetilde{V}_{WB}(z) & = &  
 - 
\left[
\frac{10 (-3+\lambda ) \lambda ^2}{N^3} \right]
J^{cd} J^{ce} K^{ef} K^{df}(z)  
-\left[ \frac{3 (-3+\lambda ) (-2+\lambda ) (-1+\lambda )}{2 N}
\right]
\pa J^{ab} \pa J^{ab}(z)  \nonu \\
& + &   \left[ \frac{(-3+\lambda ) (-2+\lambda ) (-1+\lambda
    )}{N}\right] 
\pa^2 J^{ab}  J^{ab}(z) 
 +  
  \left[ \frac{7 (-3+\lambda ) (-2+\lambda ) \lambda }{2 N} \right]
\pa^2 J^{ab}  K^{ab}(z) \nonu \\
& -&   \left[ \frac{(-3+\lambda ) \lambda  (-1+3 \lambda )}{N}
\right] \pa J^{ab} \pa K^{ab}(z) 
+
\left[ \frac{(-2+\lambda ) \lambda  (-7+29 \lambda )}{14 N} \right] J^{ab} \pa^2 K^{ab}(z) 
\nonu \\
&+ &   \left[ \frac{10 (-3+\lambda ) (-2+\lambda ) \lambda }{N^2} \right] 
J^{ab} \pa J^{ac} K^{bc}(z) + \left[ \frac{40 (-2+\lambda ) \lambda
    ^2}{7 N^2} \right]
J^{ab} K^{ac} \pa K^{bc}(z).
\label{finfinfin}
\eea
Of course, one can rewrite this (\ref{finfinfin}) 
using the equations (B.2) in
terms of quartic fields only. 
When one acts  this spin-$4$ zero mode 
on the primary states, one sees that all the $N$-dependence 
disappears and it leads to the equation (\ref{EEigeneigen}).  

\section{The fourth-order Casimir operator of $D_{\frac{N}{2}}=SO(N)$
  where $N$ is even}

In this section, the spin $4$ primary field, its 
large $N$ limit and the three-point functions
with scalars are constructed as previous section.

\subsection{ Primary spin-4 current}

The $WD_{\frac{N}{2}}$ algebra \cite{LF,LF1} is generated by the fields of spins 
\bea
2, 4, \cdots, (N-2), \frac{N}{2} \qquad N: \mbox{even}.
\label{series1}
\eea
The orders of the independent Casimir operator for the 
simple Lie algebra $D_{\frac{N}{2}}$ 
are $2, 4, \cdots, (N-2)$ and $\frac{N}{2}$. 
Since $D_2 \simeq A_1 \times A_1$ is not simple, one shall restrict to
$N \geq 3$.  
The operator product expansion of bosonic spin $\frac{N}{2}$ field
with itself provides
the bosonic fields of spin $2, 4, \cdots, (N-2)$:
\bea
\widetilde{U}_{WD}(z) \widetilde{U}_{WD}(w) & = & \frac{1}{(z-w)^N} \frac{2\widetilde{c}}{N}+
\frac{1}{(z-w)^{N-2}} 2 \widetilde{T}(w) +
\frac{1}{(z-w)^{N-3}} \pa \widetilde{T}(w)
\nonu \\
&+ & \frac{1}{(z-w)^{N-4}}  
 \left[ \widetilde{T} \widetilde{T}(w),
  \pa^2 \widetilde{T}(w), \widetilde{V}_{WD}(w) \right]
+ {\cal O}((z-w)^{-N+5}).
\label{uuwd}
\eea
 The highest spin 
field with spin-$(N-2)$ in (\ref{series1}) 
appears in the $\frac{1}{(z-w)^2}$ term.
We expect that the descendant of highest higher field of spin-$(N-2)$
should appear in the singular term $\frac{1}{(z-w)}$ in (\ref{uuwd}).
As in previous section, we would like to find the spin-$4$ primary field 
$\widetilde{V}_{WD}(z)$ which will be present in the
 $\frac{1}{(z-w)^{N-4}}$ singular term of (\ref{uuwd}).
There is no $\frac{1}{(z-w)^{N-1}}$ singular term.

The bosonic spin $\frac{N}{2}$ field
$\widetilde{U}_{WD}(z)$, that has $\frac{N+2}{2}$ terms, consists of 
\bea
\widetilde{U}_{WD}(z) & = &
 \ep^{a_1 a_2 \cdots a_N} \left[ A_0(N,k)  J^{a_1 a_2} \cdots J^{a_{N-1}
a_N}(z) + \cdots  \right. \nonu \\
&+ &  A_{\frac{N-i}{2}}(N,k)   J^{a_1 a_2} \cdots J^{a_i
  \, a_{i+1}} K^{a_{i+2} \, a_{i+3}}
\cdots K^{a_{N-1} a_N}(z) \nonu \\
&+ &   \left. \cdots + 
A_{\frac{N}{2}}(N,k)  K^{a_1 a_2} 
\cdots K^{a_{N-1} a_N}(z) \right].
\label{fermionwd}
\eea
The arbitrary coefficient functions  depend on the two integers $(N,k)$.
Then one can substitute (\ref{fermionwd}) into (\ref{uuwd}). 
We would like to focus on the $\frac{1}{(z-w)^{N-4}}$ singular terms.
From the singular term $\frac{1}{(z-w)^2}$ in the
operator product expansion $\widetilde{U}_{WD}(z) \widetilde{U}_{WD}(w)$,
the higher singular terms ($\frac{1}{(z-w)^n}$ where $n =3, 4,
\cdots, (N-2)$) can be obtained from the spin-$(N-2)$ field 
located at $\frac{1}{(z-w)^2}$ term, 
by contracting the remaining indices between the fields in the normal
ordered product.
The operator product expansion between the first term of 
(\ref{fermionwd}) with itself will lead to 
$\ep^{a_1 a_2 \cdots a_N} \ep^{a_1 a_2 \cdots b_N}  (J^{a_3 a_4} \cdots J^{a_{N-1}
a_N})( J^{b_3 b_4} \cdots J^{b_{N-1}
b_N})(w)$ after using the highest singular term in 
the operator product expansion between
the field $J^{a_1 a_2}(z)$
and the field $J^{b_1 b_2}(w)$.
Further contractions between the remaining expression will give rise to 
the lower spin field of spin $(N-3)$ by removing one current
(corresponding to the $\frac{1}{(z-w)}$-term of $J^{a_1 a_2}(z) J^{b_1 b_2}(w)$) or spin
$(N-4)$ by removing two currents (corresponding to the
$\frac{1}{(z-w)^2}$-term of  $J^{a_1 a_2}(z) J^{b_1 b_2}(w)$).  
Then the order of the original spin-$1$ fields, $(N-2)$, is reduced
to $(N-3), (N-4),
(N-5) \cdots, 4, 3, 2$ depending on the
location of singular terms. 
The fields of spins $3, 5, \cdots, (N-3)$  correspond to the
descendant fields of bosonic fields of spins $2, 4, \cdots, (N-2)$ 
of $WD_{\frac{N}{2}}$ minimal model. 
One expects that the descendant field for spin-$(N-2)$ field should
appear in the lowest singular term $\frac{1}{(z-w)}$ in (\ref{fermionwd}).

In this case, the analysis of (\ref{Spinspin4}) also holds.
Following the procedures in previous section,
one obtains the possible spin-$4$ fields 
and 
the spin-$4$ field in $WD_{\frac{N}{2}}$ minimal model is given by 
(\ref{Vtildeinter}): 
\bea
&& 
\widetilde{V}_{WD}(z)= c_3 J^{cd} J^{ef} K^{cd} K^{ef}(z)+
c_8 J^{cd} J^{ef} K^{ce} K^{df}(z)+ c_9 J^{cd} K^{ef} K^{ce} K^{df}(z)
\nonu \\ 
&& +
c_{10} K^{cd} K^{ef} K^{ce} K^{df}(z)
 +
c_{11} J^{cd} J^{cd} J^{ef} J^{ef}(z) + c_{12} J^{cd} J^{cd} J^{ef} K^{ef}(z)
 + c_{13} J^{cd} J^{cd} K^{ef} K^{ef}(z) \nonu \\
&& + 
 c_{14} J^{cd} K^{cd} K^{ef} K^{ef}(z) 
 +
c_{15} K^{cd} K^{cd} K^{ef} K^{ef}(z) +
c_{18} J^{cd} J^{ce} K^{ef} K^{df}(z) + c_{21} J^{cd} J^{ce} J^{df}
J^{ef}(z) \nonu \\
&& 
+ c_{22} J^{cd} J^{ce} J^{df}
K^{ef}(z) 
+ d_1 \pa J^{ab} \pa J^{ab}(z)  
+ d_2 \pa^2 J^{ab}  J^{ab}(z) 
+ d_3 \pa K^{ab} \pa K^{ab}(z)  
+ d_4 \pa^2 K^{ab}  K^{ab}(z) 
\nonu \\
&& + d_5
\pa^2 J^{ab}  K^{ab}(z) 
+ d_6 \pa J^{ab} \pa K^{ab}(z) 
+
d_7 J^{ab} \pa^2 K^{ab}(z) 
+ d_8 
J^{ab} \pa J^{ac} K^{bc}(z) 
\nonu \\
&& + d_9
J^{ab} K^{ac} \pa K^{bc}(z).
\label{Vtildeinteragain}
\eea
One uses the relations (B.1), (\ref{spin4derivative}) and (B.2).
Then one uses the two requirements
(\ref{regular}) and 
(\ref{tprimary}) in order to determine the coefficient functions in 
(\ref{Vtildeinteragain}).
In Appendix $G$, the requirements (\ref{regular}) and (\ref{tprimary})
are imposed and the coefficient functions appearing the spin-$4$ field
in (\ref{Vtildeinteragain}), in terms of finite $(N,k)$, are determined. 
It turns out that they are written in terms of 
two unknown coefficient functions $c_8$ and $c_{10}$.
Here we impose the following conditions for the coefficient functions
\bea
&& c_1=c_2=c_4=c_5=c_6=c_7=c_{16}=c_{17}=c_{19}=c_{20}= c_{23}=c_{24}=c_{25}=0.
\label{zerocoeff1}
\eea
Note that $c_8, c_{21}$ and $c_{22}$ are nonzero in this case.
The whole independent terms 
consists of $12$ quartic terms and $9$ derivative terms.
In order to determine the above coefficients $c_8$ and $c_{10}$, one
should compute the operator product expansion $\widetilde{V}_{WD}(z)
\widetilde{V}_{WD}(w)$.
In the Appendix $H$, we present the field contents for the $WD_3$
algebra corresponding to $N=6$ case.

The primary spin-$4$ current which is fourth order Casimir operator of
$SO(N)$ where $N$ is even is given by 
(\ref{Vtildeinteragain}) with the coeffcient functions in (G.3).
In next subsection, we describe this primary spin-$4$ current in the
large $N$ limit and find three-point functions with scalars, as in
previous analysis for $WB_{\frac{N-1}{2}}$ minimal model. 

\subsection{ Primary spin-$4$ current in the large $N$ 't Hooft limit
  and three-point functions with two scalars}

By substituting the coefficient functions (G.4), in the large $N$
limit,
into (\ref{Vtildeinteragain}) and evaluating
the correct eigenvalues, one arrives at the final contributions 
for the spin-$4$ zero mode eigenvalue
acting on the representation $( v ; 0) \otimes ( v ; 0)$, where 
$J_0^{ab} +K_0^{ab} =0$, from (G.5) and (G.6),
\bea
\widetilde{V}_0 | {\cal{O}}_{+} > & = & \left[
-\frac{c_{8} N^2 \left(-1+2 \lambda +39\lambda ^2\right)}{4
  \lambda ^2}+\frac{2 c_{10} N^4 \left(6+11 \lambda +56 \lambda
    ^2+11 \lambda ^3\right)}{5 \lambda ^4} \right]
| {\cal{O}}_{+} >,
\label{equ1}
\eea
where
$ {\cal{O}}_{+} \equiv 
( v ; 0) \otimes ( v ; 0)$ which is equivalent to $ (2,
1^{\frac{N}{2}-1}|1^{\frac{N}{2}}) 
\otimes (2,
1^{\frac{N}{2}-1}|1^{\frac{N}{2}}) $ in the
convention of \cite{Ahn1106}.

For the second primary, one has vanishing $K_0^{ab}$ and this implies
that there exist contributions from the $c_{11}$-, $c_{21}$-, $d_1$- and
$d_2$-terms.
The leading contribution $N^4$ from $c_{10}$ factor comes from the coeffcient
functions,
$d_1$ and $d_2$,
leads to
\bea
&& - 2N d_1  -4N  d_2  \rightarrow
- \left[ \frac{2 
  N^4 (-3+\lambda ) (-2+\lambda ) (-1+\lambda )}{5 \lambda ^4} \right]
c_{10},
\label{resres}
\eea
and the leading contribution $N^2$ from $c_{8}$ factor comes from the coeffcient
functions,
$c_{11}$, $c_{12}$, $d_1$ and $d_2$,
leads to
\bea
&& 4N^2 c_{11} + N^2 c_{21} - 2N d_1  -4N  d_2  
\rightarrow
\left[ \frac{  N^2 (-1+\lambda )^2}{4 \lambda ^2} \right] c_8.
\label{resres1}
\eea
Using (\ref{11-15dep}), (\ref{21-25dep}) and (\ref{quadraticdep}) with
correct multiplicities for the Fourier mode on the derivative terms,
the following spin-$4$ zero mode eigenvalue equation reads, from (\ref{resres})
and (\ref{resres1}),
\bea
\widetilde{V}_0 | {\cal{O}}_{-} > & = & 
\left[ \frac{ c_{8} N^2 (-1+\lambda )^2}{4 \lambda ^2}-\frac{2 c_{10}
  N^4 (-3+\lambda ) (-2+\lambda ) (-1+\lambda )}{5 \lambda ^4} \right]
 | {\cal{O}}_{-} >,
\label{equ2}
\eea
where
$ {\cal{O}}_{-} \equiv 
( 0; v ) \otimes ( 0; v )$ which is equivalent to $ (1^{\frac{N}{2}}|2,
1^{\frac{N}{2}-1}) 
\otimes (1^{\frac{N}{2}}|2,
1^{\frac{N}{2}-1}) $ in the
convention of \cite{Ahn1106}.

In this case, the spin-$2$ Virasoro zero mode eigenvalues are fixed by
the conformal dimension as before, (E.10) and (E.12). 
Moreover, the three-point functions with scalars are given by
(E.13).
Once again, from the observation of \cite{GV},
the eigenvalues are given by $(1\pm \la)(2\pm \la)(3 \pm \la)$
on the primaries $|{\cal O}_{\pm}>$.
For the choice of
\bea
c_{8}(N,\la) & = & \frac{20 (-3+\lambda ) (-2+\lambda ) \lambda ^3
  (5+\lambda )}{N^2 \left(11+134 \lambda -119 \lambda ^2+14 \lambda
    ^3\right)}, \nonu \\
c_{10}(N,\la) & = & \frac{5 \lambda ^4 \left(11+109 \lambda
    -99 \lambda ^2+19 \lambda ^3\right)}{2 N^4 \left(11+134 \lambda
    -119 \lambda ^2+14 \lambda ^3\right)},
\label{twoconstant1}
\eea
one obtains the eigenvalue equations given by (\ref{EEigeneigen})
where the two primaries are given by the above  $|{\cal O}_{\pm}>$,
from (\ref{equ1}) and (\ref{equ2}).
Finally, the three-point functions are summarized by (\ref{three}).
The undetermined two coefficient
functions $c_8(N,k)$ and $c_{10}(N,k)$ occur in this operator product
expansion $\widetilde{V}_{WD}(z) \widetilde{V}_{WD}(w)$. 
Only after this computation which will be very complicated
(i.e., $21 \times 21 = 441$ operator product expansions one should compute) is done, 
they are fixed completely. 
They should take the form
(\ref{twoconstant1}) as one takes the large $N$ limit. 

\section{Conclusions and outlook }

We have found the coset primary spin-$4$ field (\ref{Vtilde}) with
(E.7), where two coefficient functions are not fixed, 
in the $WB_{\frac{N-1}{2}}$ minimal model. 
These coefficient functions can be fixed, in principle, only after 
the $324$ operator product expansions are computed.
With appropriate choice for these coefficient functions (recalling the
higher spin Lie algebra), we have constructed the three-point
functions with two scalars in (\ref{three}) under the large $N$ 't
Hooft limit.
Furthermore, we also have described the coset primary spin-$4$ field 
(\ref{Vtildeinteragain}) 
with (G.3) in the $WD_{\frac{N}{2}}$ minimal model (with two
unknown coefficient functions) and 
found the three-point functions with scalars in the large $N$ limit
under the similar assumption on the higher spin Lie algebra.
The explicit forms for the spin-$4$ fields in the large $N$ limit 
are given in (\ref{finfinfin})
and 
(G.8).
For $WA_{N-1}$ minimal model, since all the coefficient functions
are fixed, the eigenvalue equations lead to those for higher spin
algebra automatically.
However, for $WB_{\frac{N-1}{2}}$ and $WD_{\frac{N}{2}}$ minimal models, 
we require that the eigenvalue equations should  satisfy the higher
spin Lie algebra in order to fix the undetermined coefficient
functions and after
that all the coefficient functions are determined completely.  
The complete expression for the primary spin-$4$ field with finite
$(N,k)$
is known only for the $WA_{N-1}$  minimal model so far. 
In order to obtain those for the  $WB_{\frac{N-1}{2}}$ and
$WD_{\frac{N}{2}}$ 
minimal models, one should compute the operator product expansions
explicitly as one described before.

It is simple to ask what the corresponding three-point functions in
three-dimensional higher spin gravity for the present minimal models are.
Based on the works of \cite{CY} or more recently \cite{AKP},  
it is an open problem to compute the three-point functions in the bulk
for any
deformation parameter $\la$.

In this paper, we have considered only higher spin field of fixed spin
$s=4$.
According to the observation of \cite{AKP}, the three-point functions
are written for arbitrary spin $s$.
Via the AdS/CFT duality in \cite{GG}, one should see those three-point
functions in the $W_N$ minimal model conformal field theory in the
large $N$ limit. This implies that the results of \cite{Ahn1111} and
the present paper should be generalized to the construction of coset
Casimir operators of arbitrary spin $s$.
It would be interesting to find the Casimir operators of spin $s$ in
the $W A_{N-1}, WB_{\frac{N-1}{2}}$, and $WD_{\frac{N}{2}}$ minimal models.

The two undetermined coefficient functions in the present minimal
models
cannot be fixed 
by the requirements that it should be a primary field of spin-$4$
with respect to the spin-$2$ coset Virasoro field and that 
it should commute with the diagonal subalgebra.
Without computing the operator product expansions of spin-$4$ field
with itself, are there any ways to compute 
the unknown two coefficient functions explicitly?
If one considers the extended ${\cal N}=1$ supersymmetric algebra
which contains the field contents we have discussed in this paper and
its superpartners, one can construct the spin-$\frac{3}{2}$ field
$\widetilde{G}(z)$ which is a fermionic partner of coset spin-$2$
Virasoro field $\widetilde{T}(z)$ 
in
the $WB_{\frac{N-1}{2}}$ minimal model, along the line of \cite{ASS,BS}. 
For the $WD_{\frac{N}{2}}$ minimal model, it is not clear how to
construct odd (fermionic) spin current. 
Then one can compute the operator product expansion between
$\widetilde{G}(z)$
and spin-$4$ field $\widetilde{V}_{WB}(w)$.
In the right hand side of this operator product expansion, 
one expects that the highest singular term $\frac{1}{(z-w)^4}$ should
be proportional to $\widetilde{G}(w)$. Then this will determine the
unknown coefficient functions under the above assumption.
It would be interesting to find whether the $WB_{\frac{N-1}{2}}$
algebra can be extended to the extended ${\cal N}=1$ superconformal
algebra or not.     
 
\vspace{.7cm}

\centerline{\bf Acknowledgments}

This work was supported by the Mid-career Researcher Program through
the National Research Foundation of Korea (NRF) grant 
funded by the Korean government (MEST) (No. 2009-0084601).
CA acknowledges warm hospitality from 
the School of  Liberal Arts (and Institute of Convergence Fundamental
Studies), 
Seoul National University of Science and Technology.

\newpage


\end{document}